\documentclass{PoS}

\usepackage{graphicx}

\title{IGR J17091-3624: a GRS 1915-105 like source as seen by INTEGRAL and Swift}

\ShortTitle{IGR J17091-3624: a GRS 1915-105 like source as seen by INTEGRAL and Swift}

\author{\speaker{Fiamma Capitanio}~\thanks{FC, MDS, AP and GDC acknowledge financial support from the
agreement ASI-INAF I/009/10/0}\\
        INAF/IAPS Rome, Italy\\
        E-mail: \email{fiamma.capitanio@iaps.inaf.it}}

\author{Melania Del Santo~\thanks{MDS and GDC  acknowledge financial contribution by the grant
 PRIN-INAF 2009 (PI: L. Sidoli)}\\
INAF/IAPS Rome Italy        \\
}
\author{Enrico Bozzo\\
ISDC Geneva  Switzerland \\
}
\author{Carlo Ferrigno\\
ISDC Geneva  Switzerland       \\
}
\author{Giovanni De Cesare\\
INAF/IAPS Rome Italy   \\
}
\author{Adamantia Paizis\\
INAF/IASF Milan Italy   \\
}
\abstract{
We present here the main characteristics of the BHC IGR 
J17091-3624 outbursts occurred several times since 1994. Since 2003, the source has been extensively 
observed  by INTEGRAL and Swift. In particular, we report results  on the last 2011 outburst that showed a rare 
variability behaviour observed before only in the galactic BH GRS 1915+105 but at a different level of flux. 
Several hypotheses have been proposed in order to explain this particular behaviour. They are all discussed here, 
in the light of their apparent contradiction. Finally, based on all available informations, we attempt to give an overall 
view of this enigmatic source and we speculate on the evolutionary state of the binary system.
}

\FullConference{ An INTEGRAL view of the high-energy sky (the first 10 years) - 9th INTEGRAL Workshop and celebration of the 10th anniversary of the launch\\
                 15-19 October 2012\\
                 Bibliotheque Nationale de France, Paris, France}
\begin{document}

\section{Introduction}
\begin{figure}[h!] %  figure placement: here, top, bottom, or page
   \centering
   \includegraphics[scale=0.55,angle=90]{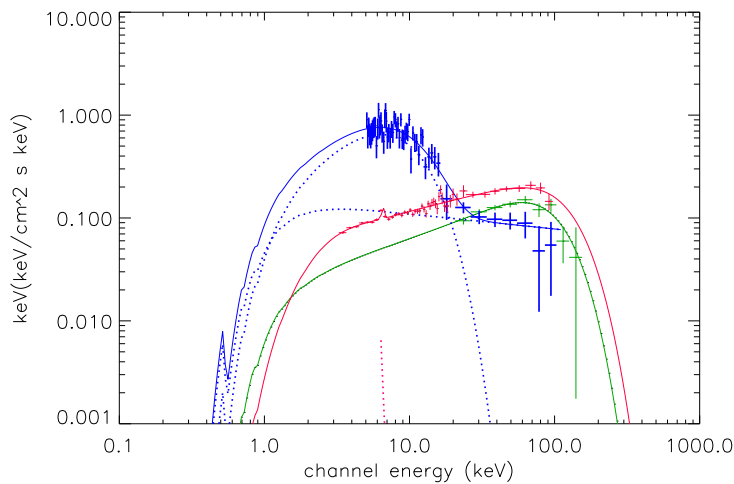} 
   \caption{Three spectra of the 2003 outburst, from Capitanio et al.(2006). Red spectrum: INTEGRAL/IBIS and RXTE/PCA spectrum during the hard state at the beginning of the outburst. Blue spectrum: INTEGRAL IBIS and JEM-X spectrum during the soft state. Green spectrum: INTEGRAL/IBIS spectrum during hard state at the end of the outburst.}
   \label{2005spec}
\end{figure}

IGR J17091--3624 was discovered by INTEGRAL on 2003~\cite{Kuulkers}. %and classified as Black Hole candidate thanks to its spectral behaviour during the outbursts. 
Subsequently the source was searched (and found) in the data archives of other missions with the result that from 1994 until 2011, IGR J17091--3624 appeared as a moderately bright transient source, at about 10-20 mCrab (20-100 keV) flux level, with flaring activity in: \\
%\begin{itemize}
%\item
1) 1994 October (Mir-Kvant  TTM)~\cite{Zand}; \\
%\item 
2) 1996 September (BeppoSAX WFC)~\cite{Zand};\\
%\item 
3) 2001 September (BeppoSAX WFC)~\cite{Zand};\\
%\item
 4) 2003 April (INTEGRAL IBIS)~\cite{Kuulkers};\\
%\item 
5) 2007 July (Swift/XRT)~\cite{Kennea}; \\
%\item 
6) 2011 February (Swift/BAT)~\cite{Krimm}.\\
%\end{itemize}
All the outbursts of IGR J17091--3624 observed before 2011, in the limit of the instruments capabilities, displayed the typical spectral and temporal evolution 
as expected from a standard Black Hole Candidate (BHC) ~\cite{Cap1,Cap2}. A transient radio counterpart was  found in the archival data of the VLA  (8.4 GHz) during the 2003 outburst.
Figure~\ref{2005spec} shows three different spectra observed by INTEGRAL and RXTE during the 2003 outburst.

\begin{figure}[h!] %  figure placement: here, top, bottom, or page
  \centering
   \includegraphics[scale=0.7]{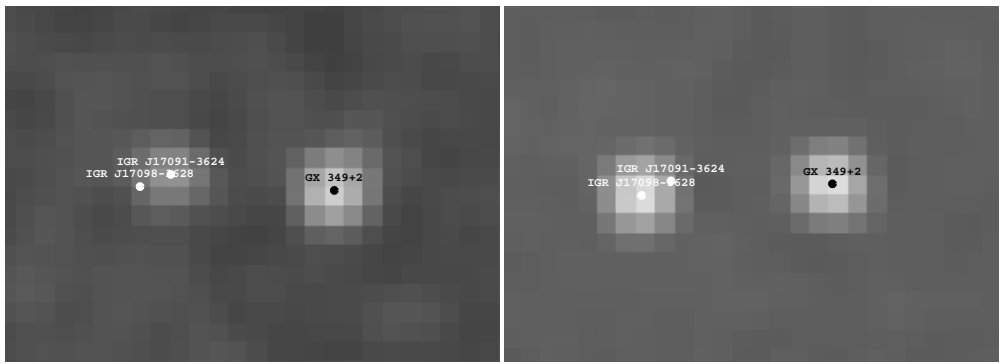} 
   \caption{Left: 2003 IBIS/INTEGRAL 80 ks mosaic image (20--40 keV) of IGR J17091--3624 and IGR J17098--3628 zone: only IGR J17091--3624 is visible. Right:
2005 IBIS/INTEGRAL 120 ks mosaic image (20--40 keV) of IGR J17091--3624 and IGR J17098--3628 zone: only IGR J17098--3628 is visible. Image from Capitanio et al. (2009)}
   \label{image}
\end{figure}

IGR J17091-3624  underwent again in outburst in July 2007. The detection of the source was particularly difficult during this period because of the bright outburst of the nearby transient source IGR J17098--3628 that had started on March 2005. In fact, the two transient sources lie at 9.6 arcmin away from each other, thus,  while for example INTEGRAL/IBIS is able to distinguish which of the two sources is active (see Figure~\ref{image}  and  Grebenev et al. 2007 for details), other X-ray wide field monitors, such as RXTE/ASM or Swift/BAT, could not distinguish between them~\cite{Kennea}. 
%
%between the two sources when only one is detectable,  (see Figure~\ref{image}  and  Grebenev et al. 2007 for details), the other X-ray monitors, such as RXTE/ASM and the Swift/BAT, could not distinguish if the emission comes from IGR J17091--3624 or IGR J17098--3628~\cite{Kennea}.  %In fact the ASM light curve of IGR J17091-3624 is the sum of the emission of the two sources.
%Figure~\ref{image} shows two INTEGRAL/IBIS mosaic images of the IGR J17091-3624 and IGR J17098-3628 zone during 2003 and 2005 outbursts, when only one of the two sources was detected.
\section{The 2011 outburst, a special case}

The 2011 outburst was the brightest ever observed for IGR J17091-3624 (with a peak flux of about 100 mCrab between 20-40 keV). The source activity was monitored by a long XRT ToO follow-up monitor campaign from  January until August 2011.
Whenever possible, the XRT observations were performed simultaneously with INTEGRAL. A total of 72 Swift/XRT pointings were collected, 19 of which simultaneously with INTEGRAL. 
The beginning of the outburst was similar to the previous ones: during the initial hard state the source flux increased with essentially a constant spectrum~\cite{Cap3}. The spectrum is well fitted with an absorbed power law with a cut-off at 104$\pm$14 keV and a photon index of 1.5$\pm$0.1~\cite{Cap3, Rodriguez}. Otherwise, considering a simple Comptonization model such as the {\it Comptt} model~\cite{Tit}, we find that the electron temperature goes from  22 until 30 keV while the optical depth is in the range of 1.6-2.0. These results are consistent with the ones reported by Capitanio et al. 2006 for the 2003 IGR J17091-3624 outburst.   

\begin{figure}[h!] %  figure placement: here, top, bottom, or page
   \centering
   \includegraphics[scale=0.45,angle=90]{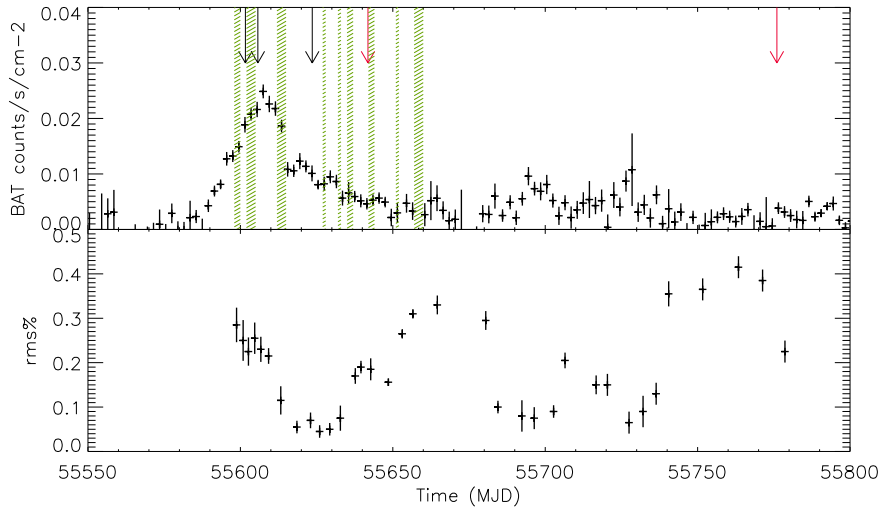} 
   \caption{Top panel: Swift/BAT 15-50 keV light curve, the black arrows represent the period of the radio observations in which the source is detected (from Rodriguez et al. 2011). The red arrows represent the periods in which the source is not detected in the radio band (from Rodriguez et al. 2011 and King et al. 2012). The green parts represent the INTEGRAL observation periods.The bottom panel shows the corresponded {\it rms} values as function of time.}
   \label{Figure3}
\end{figure}

\begin{figure}[h!] %  figure placement: here, top, bottom, or page
   \centering
   \includegraphics[scale=0.4,angle=-90]{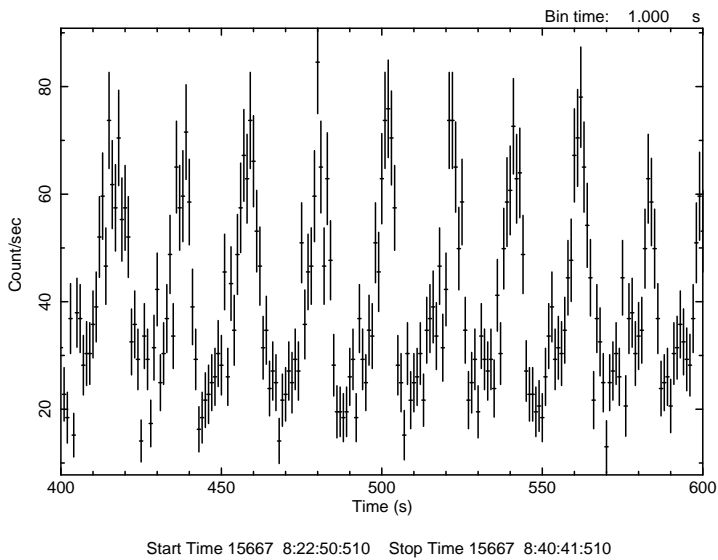} 
   \caption{Zoom of the 0.3-10 keV XRT light curve of IGR J17091-3624. XRT observation num. 00031921042. Bin time = 1s}
   \label{Figure4}
\end{figure}

On MJD 55614, the source displayed evidence of a spectral transition to the soft state~\cite{Rodriguez,Cap3}: the flux continued to increase rapidly and a significant softening of the hard X-ray spectrum was observed, together with a drop in the hard X-ray flux. Then, as Figure~\ref{Figure3} shows, the  {\it rms} started to decrease and, in the same period,  a radio flare was reported by Rodriguez et al. 2011. Figure~\ref{Figure3} shows the Swift/BAT 15-50 keV light curve (top panel), the black arrows represent the period of the radio observations in which the source is detected (from Rodriguez et al. 2011), while the red arrows represent the periods in which the source is not detected in the radio band (from Rodriguez et al. 2011 and King et al. 2012). The bottom panel shows the corresponded {\it rms} values as function of time. As we can see in Figure 3, after the initial decrease, due to the transition to the soft state, the {\it rms} value increases again and then it starts to fluctuate.
 As reported by Capitanio et al. 2011, the higher values of the {\it rms} correspond to the presence of a quasi periodical flare like events that resemble the GRS 1915+105 so  called "{\it Heartbeat}" (see Figure~\ref{Figure4}). 
The presence of such oscillations in the light curve were reported for the first time by Altamirano et al. 2011  and have not ever been observed before in the XRT data of the previous outbursts~\cite{Cap3}. The "{\it Heartbeat}" behaviour and evolution are similar for the two sources ~\cite{Cap3, Neilsen, Mineo}.

%The IGR J17091-3624 spectral states resemble the ones of GRS 1915+105 (Belloni et al 1997) as reported by Altamirano et al 2011.

The "{\it Heartbeat}" phenomenon is probably due to the Lightman-Eardley instability~\cite{Lightman}, a limit cycle in the inner accretion disc dominated by the radiation pressure.
The inner part of the disc empties and refills with a time-scale of seconds.  This instability is expected to occur when the luminosity of the system is comparable to the Eddington one~\cite{Ohsuga}.

For IGR J17091--3624 we do not have an estimation of the distance, the inclination angle, the spin, the BH mass, and the characteristics of the companion star. Rodriguez et al. (2011) and  Pahari et al. (2011) estimated the source distance (at the beginning of the outburst): 11-20~kpc, Pahari et al. (2011) estimated also the BH mass range: M=8-11~M$_{\odot}$.
Considering these parameters and the observation with the highest flux that displays the "{\it Heartbeat}", the luminosity of the source spans from 1-8\%L$_{edd}$~\cite{Cap3} .
Since the flare-like events should be at Eddington limit regime (see e.g. Neilsen et al. 2011), the faintness of IGR J17091--3624 should not be only due to the source distance. 
Several hypotheses have been proposed in order to explain the faintness of the IGR J17091-3624 "{\it Heartbeat}".
For example, Capitanio et al. 2011 speculate that the lower luminosity of IGR J17091--3624 could also be ascribed to the spectral deformation effects due to the high inclination angle that is favored by the data~\cite{King,Cap3,Rao}. 
In fact a study, reported by Cunningham (1975), claims that when a Kerr BH is seen at a high inclination angle (cosi $<$ 0.25, i $>$75$\deg$), the source appears significantly fainter (up to a factor that depends on the BH spin and mass but can reach about an order of magnitude less) with respect to the system observed face-on. At odds with this hypothesis is the lack of detection of eclipses that could be related to a small ratio between the companion star and the BH mass~\cite{Eggleton}. 
Other authors report different conclusions that we briefly  summarize below:
\begin{itemize}
\item Altamirano et al. 2011  calculated that if the "{\it Heartbeat}" emission (as for GRS 1915+105) requires near Eddington luminosities, IGR J17091--3624 should lie at a distance greater than 20 kpc or harbor one of the least massive black holes known (< 3M$_{\odot}$).
\item Recently P. Rebusco et al. 2012, from HFQPOs, estimated a BH mass of 6M$_{\odot}$, for which the distance extrapolated from Altamirano et al. 2011 should be $>$ 22 kpc, thus approximately outside the Galaxy.
\item Rao \& Vadawale 2012 try to explain the faintness of IGR J17091--3624, from spectral analysis,  considering a low or even retrograde spin together with an high inclined system. 

\item A recently published Astronomical Telegram~\cite{Reis}, confutes this hypotheses claiming that (from XMM data) the spin of the source should be very high (a $\sim$1) and the inclination lower than 62~$\deg$.

\end{itemize}  
\section{Conclusions}

Some of the results, reported in the previous section, are  in contradiction with each other. However due to the lack of any information about the system parameters,  we cannot accept or reject a priori any of them. Finally in the following paragraphs we try to list some open questions related to this enigmatic source that are a challege for the future multiwavelenght observations:
 
\begin{itemize}

\item IGR J17091--3624 can no longer be considered as a typical Black Hole transient. In fact, after the transition from the hard to the soft state~\cite{Rodriguez}, the source did not follow the standard q-track in the Hardness Intensity Diagram (HID)~\cite{Homan} and, since March 2011, it remained trapped in the top left corner of the HID (for details see Capitanio et al. 2012) %as Figure~\ref{HID} shows
~\cite{Cap3}.
As of the date of the present work, the source reached about two years of activity. In spite of the long duration,  it is  not the longest outburst ever seen from a transient BHC (see for example the nearby X-ray transient IGR J17098-3628~\cite{Kotze}).  {\it Are these changes in IGR J17091-3624 behaviour representing an evolution from a transient source to a persistent one as happened for GRS 1915+105?}
\item Another peculiarity of this source is the presence of a particular fast and ionized wind (anti-correlated with jet), observed during the  "{\it Heartbeat}" soft spectral state, with a velocity of about 9000 km/s ~\cite{King}. {\it Is this extreme fast wind correlated with the faint pseudo periodical oscillations of the light curve?}
%\item IGR J17091-3624 is an GRS 1915+105 like source. In fact it presents not only the "heartbeat" behaviour but also other spectral states that resemble the one observed for GRS 1915+105~\cite{Altamirano}. However all these similarities do not explain the faintness of the source
\item In the light of all of the hypotheses reported in the previous section, {\it can we speculate that  the IGR J17091-3624 "{\it Heartbeat}"  are possible even at luminosity lower than the Eddington one?~\cite{Janiuk}}
\end{itemize}
%\begin{figure}[h!] %  figure placement: here, top, bottom, or page
   %\centering
   %\includegraphics[scale=0.35,angle=0]{Figure5.ps} 
  % \caption{Hardness Intensity Diagram of all the XRT 2011 outburst observations of IGR J17091--3624.}
   %\label{HID}
%\end{figure}

%The X-ray flux intensity of the heartbeat  
%	20-40  keV averaged flux (from Bird et al. 2010): 
%	GRS 1915+105: 284 mCrab; 
%	IGR J17091-3624: 2011 20-40 keV peak flux ~100 mCrab
%Only refined estimation of the distance and the BH mass might help to answer to these questions.

\end{document}